\documentclass[a4paper,twoside,10pt]{letter}
\usepackage{graphicx,saj,multicol,subeqnarray}

\def\papertype{{\doi DOI: 10.2298/SAJ1591001C}\hfill\ Invited review}

\def\udc{57:52 $+$ 573.5}
\begin{document}
\baselineskip=3.1truemm
\columnsep=.5truecm
\newenvironment{lefteqnarray}{\arraycolsep=0pt\begin{eqnarray}}
{\end{eqnarray}\protect\aftergroup\ignorespaces}
\newenvironment{lefteqnarray*}{\arraycolsep=0pt\begin{eqnarray*}}
{\end{eqnarray*}\protect\aftergroup\ignorespaces}
\newenvironment{leftsubeqnarray}{\arraycolsep=0pt\begin{subeqnarray}}
{\end{subeqnarray}\protect\aftergroup\ignorespaces}
%

% Running titles

\markboth{\eightrm KARDASHEV'S CLASSIFICATION AT 50+}
{\eightrm M. M. \'CIRKOVI\'C}

{\ }

%\publ

%\type

{\ }

% Title

\title{KARDASHEV'S CLASSIFICATION AT 50+:\break A FINE VEHICLE WITH ROOM FOR IMPROVEMENT}

% Authors

\authors{M. M. \'Cirkovi\'c$^{1,2}$}

\vskip3mm

% Address

\address{$^1$Astronomical Observatory, Volgina 7, 11060 Belgrade 38, Serbia}

\address{$^2$Future of Humanity Institute, Faculty of Philosophy, University of Oxford,\break
Suite 8, Littlegate House, 16/17 St Ebbe's Street, Oxford, OX1 1PT, UK}

\Email{mcirkovic@aob.rs}

% Received and Accepted dates

\dates{November 27, 2015}{November 27, 2015}

% Abstract

\summary{We review the history and status of the famous
classification of extraterrestrial civilizations given by the
great Russian astrophysicist Nikolai Semenovich Kardashev,
roughly half a century after it has been proposed. While
Kardashev's classification (or Kardashev's scale) has often been
seen as oversimplified, and multiple improvements, refinements,
and alternatives to it have been suggested, it is still one of
the major tools for serious theoretical investigation of SETI
issues. During these 50+ years, several attempts at modifying or
reforming the classification have been made; we review some of
them here, together with presenting some of the scenarios which
present difficulties to the standard version. Recent results in
both theoretical and observational SETI studies, especially the
\^G infrared survey (2014-2015), have persuasively shown that the
emphasis on detectability inherent in Kardashev's classification
obtains new significance and freshness. Several new movements and
conceptual frameworks, such as the Dysonian SETI, tally extremely
well with these developments. So, the apparent simplicity of the
classification is highly deceptive: Kardashev's work offers a
wealth of still insufficiently studied methodological and
epistemological ramifications and it remains, in both letter and
spirit, perhaps the worthiest legacy of the SETI ``founding
fathers''.}


\begin{multicols}{2}
{

{\ }

\vskip.5cm

\hfill\ \textit{By their fruits ye shall know them.}

\vskip2mm

\hfill\ Matthew, 7:16

\vskip6mm

\hfill\ \textit{Once out of nature I shall never take}

\hfill\ \textit{My bodily form from any natural thing...}

\vskip2mm

\hfill\ William Butler Yeats

\vskip.5cm

\section{1. INTRODUCTION: KARDASHEV'S LADDER}

One of the achievements of the early days of the Search for ExtraTerrestrial
Intelligence (SETI), more than a half century ago now, was a practical way
of thinking how to classify potential search targets by their impact on
physical environment. The expression of this was the famous Kardashev's
classification (or Kardashev's scale) of advanced extraterrestrial
societies, originally containing three types of civilizations detectable, at
least in principle, through practical SETI
activities:\footnote{Kardashev
(1964), p. 219. I shall use Arabic numerals for Kardashev's types throughout
this study, although it is a historical fact that Kardashev, and indeed most
subsequent authors, used Roman numerals. Apart from the latter being
outdated in general, I have two justifications specific to the problem at
hand: (i) Arabic numeration enables natural introduction of fractional
subtypes, like Type 2.5 civilization, etc. which was already a problem for
Carl Sagan in 1970s -- see the quote below; and (ii) there is no Roman
numeral for zero, while it seems logically natural to introduce Type 0
civilization (and its fractional successors) as the pre-technological state
of any intelligent community. More on this in Section 2.}

\mc{\textit{[I]t will prove convenient to classify technologically
developed civilizations in three types:}

\vskip2mm

\noindent\textit{I -- technological level close to the level
presently attained on the earth, with energy consumption at $
\approx $ 4 $\times $ 10$^{19}$ erg/sec.}

\vskip2mm

\noindent\textit{II -- a civilization capable of harnessing the energy
radiated by its own star (for example, the stage of successful construction
of a ``Dyson sphere''...); energy consumption at $ \approx $ 4 $\times $
10$^{33}$ erg/sec.}

\vskip2mm

\noindent\textit{III -- a civilization in possession of energy on the scale
of its own galaxy, with energy consumption at $ \approx $ 4 $\times $
10$^{44}$ erg/sec.}}

In other words, and in more conventional reading, we are dealing with the
following basic types:\footnote{See Shklovskii and Sagan (1966) as the
``urtext`` in this respect; Michaud (2007) or Bennett and Shostak (2011) for
the prototypical ``textbook'' approach to the classification (and indeed
most SETI issues).}

\vskip2mm

\textsc{\textbf{Type 1:}} a civilization manipulating energy
resources of its home planet.

\textsc{\textbf{Type 2:}} a civilization manipulating energy resources of
its home star/planetary system.

\textsc{\textbf{Type 3:}} a civilization manipulating energy resources of
its home galaxy.

\vskip2mm

Why is taxonomy important? Claude L\'{e}vi-Strauss famously
argued that ``Darwin would not have been possible if he had not
been preceded by Linnaeus.'' Historical experience in many other
fields of science (chemistry, particle physics, extragalactic
astronomy) strongly confirms this dictum. The underlying idea is
that the very act of formulating explanatory hypotheses in any
field is impossible to perform without an appropriate taxonomical
framework. And the historical fact that Linnaeus held views
about, say, biological species and their origination and
persistence which were \textit{diametrally opposed} to what we
regard as basic tenets of the Darwinian revolution does not
influence L\'{e}vi-Strauss' conclusion in the least. Linnaeus'
personal beliefs about the origin of species and other taxons were
unimportant; his taxonomy was the necessary, indeed magic, key to
understanding. This could be immediately applied to the SETI
research as well: without prejudicating anything about the
outcome of the SETI searches or indeed our theoretical views on
the emergence and frequency of extraterrestrial civilizations, we
still need a taxonomical scheme indicative of \textit{what we
might expect to find}. Our personal beliefs about the existence
of SETI targets and the likelihood of success in the entire
endeavor are unimportant; it is the taxonomy that matters. Since
hypothetical targets of any particular SETI programme are likely
to be wildly non-uniform -- due to contingency of biological
evolution, if nothing else (Gould 1989, Conway Morris 2003, 2011)
-- it is only reasonable that such taxonomic scheme is rather
coarse-grained. In addition, it should take into account the
realistic discrete distribution of matter -- in essence, the fact
that the distribution of matter (both baryonic and non-baryonic)
in the universe is distinctly clumpy and that the emergent complex
configurations of matter including living and intelligent and
technological systems need to follow that same clumpiness.
Kardashev's Types thus correspond to the most important (from the
point of view of any practical search) classes of objects we
encounter in the universe, namely planets, stars (with their
planetary systems) and galaxies. While minor deviations from the
principle that life and intelligence follows the distribution of
matter are possible -- and indeed desirable -- when considering
advanced technological societies or taking into account the
possibility of interstellar panspermia, these baseline celestial
bodies remain the foci of any observational and theoretical
search.

Practical \textit{need for a taxonomy\/} in SETI studies and the
\textit{hierarchical distribution of matter\/} are two legs on
which the significance of Kardashev's scale rests; the other two
are \textit{Copernicanism\/} and \textit{universality of
evolution}. Copernicanism (often called the Principle of
Mediocrity, the Principle of Typicality, etc.) in the narrow
sense tells us that there is nothing special about the Earth or
the Solar System or our Galaxy within large sets of similar
objects throughout the universe. In somewhat broader sense, it
indicates that there is nothing particularly special about us as
observers: our temporal or spatial location, or our location in
other abstract spaces of physical, chemical, biological, etc.,
parameters are typical or close to typical. Copernicanism did not
only played an important role in the great scientific revolution
which coined the moniker, but continues to play a vital part in
debates surrounding both classical and quantum cosmology, and in
particular attempts to apply various overarching theories of
fundamental physics to an ensemble of universes, or the
multiverse (e.g. Ellis et al. 2004, Page 2008, Linde and
Vanchurin 2010). In the specific case of emerging SETI theory, we
have been witnessing attempts to use Copernicanism in order to
construct models of the set of habitable planets in the Galaxy
(Franck et al. 2007, Vukoti\'c and \'Cirkovi\'c 2012, Hair
and~Hedman 2013) or to constrain the evolution of intelligent
observers (Carter 2008, 2012, Gleiser 2010). Kardashev's
classification relies on Copernicanism for the underlying
assumption that both the increase in energy consumption and the
overall resources in the astrophysical environment are in general
typical for the universe at large; some advanced exceptions to
this will be considered below.

\vskip.7cm

\centerline{\includegraphics[width=7cm]{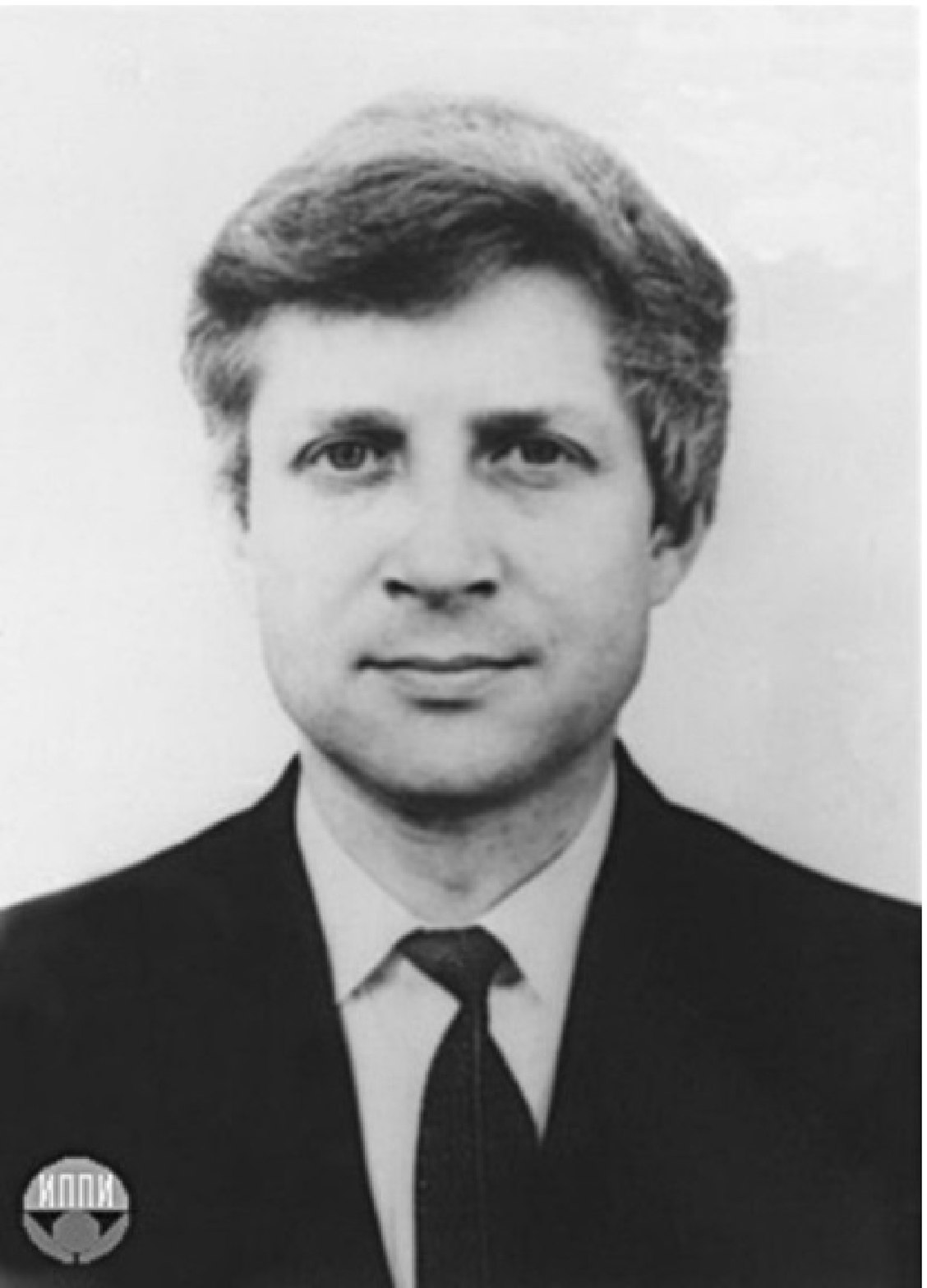}}
\label{fig1}
\figurecaption{1.}{Academician Nikolay Semenovich
Kardashev, the official photo. $\!$ (\textit{Courtesy of the Russian
Academy of Sciences.})}

The \textit{evolutionary character\/} of the classification was
rather obvious at the time of its origin, amidst all the great
excitement and enthusiasm for SETI in 1960s and 1970s. Kardashev
and the rest of the ``founding fathers'' (Drake, Morrison,
Bracewell, Oliver, Cocconi, Papagiannis, Shklovsky, and Sagan)
clearly perceived SETI as a means for verifying the assumptions
about biological \textit{and }cultural evolution they deemed
``natural'' or ``typical'' or ``default''. Hence came a rather
violent reaction of some critics for perceived trampling on their
hallowed turf of inquiry, either biological (e.g. G. G. Simpson,
E. Mayr) or philosophical (e.g. N. Rescher, E. McMullin, A.
Kukla).\footnote{Simpson (1964), Rescher (1985), Mayr (1993),
Kukla (2001). Simpson's strong-worded attack on SETI is the
prototype of this sort of criticism. See, however, \'Cirkovi\'c
(2014) for the possibility of more fruitful reinterpretation of
Simpson's criticism with today's hindsight and in the
contemporary astrobiological context.} This reaction could not,
however, turn the wheel of history backward: the role of mind and
intelligence in the universe at large -- what was for a long time
the province of a few bold speculative and mystical authors like
Tsiolkovsky or H. G. Wells or Stapledon -- has become part of the
scientific discourse (for the historical accounts see Crow 1986,
Dick 1996, Kragh 2004). In this context, the emergence of
Kardashev's scale as a practical ``rule of thumb'' for
quantifying this central issue -- the impact of intelligence on
the physical universe -- could be regarded as somewhat symbolical
for new directions in thinking which came before their time and
are only now reaching fruition.

So, why is Kardashev's classification still of vital importance to us after
more than 50 years of (so far unsuccessful) SETI efforts? Answers to this
question are multifold. Since 1995, we are in the period which is more and
more often referred to as ``astrobiological revolution'' (e.g. Des Marais
and Walter 1999, Grinspoon 2003, Gilmour and Sephton 2004, Chyba and Hand
2005). Rapid increase of our knowledge about the cosmic context of
abiogenesis and evolution, as well as realization that there are numerous
potential habitats for life in the Galaxy, have been accompanied by the
increase in both public interest and institutional framework, including new
research departments, new peer-reviewed journals, etc. Entirely new key
concepts, such as the \textit{Galactic Habitable Zone} (henceforth GHZ,
Lineweaver 2001, Gonzalez, Brownlee,
and Ward 2001, Gonzalez 2005), have been introduced in this period, and the
wider synergy between various fields of astronomical and life sciences has
been achieved within this wide astrobiological front.

And yet, much older SETI research (starting with Project OZMA in
1960, or in late 1950s with the work of Cocconi and Morrison
1959) has not been entirely and happily integrated into the
emerging astrobiological paradigm -- for multiple reasons, some
of which go way beyond the realm of science. On the purely
cognitive level, the need for smooth integration is obvious, since
it follows the physicalist and evolutionary foundations of all
life sciences and technology. SETI studies cannot be anything but
a particular research sector of the overall astrobiological
effort; however, there have been unhealthy tensions on both
sides, from the high level of philosophical approaches (best
manifested in the rise of the ``rare Earth'' hypothesis) down to
the overt funding issues and controversies (Darling 2001, Ward
2005). Therefore, it is of much current methodological \textit{and
practical} interest to seek those ideas and concepts which could
facilitate this integration and enable stable and fruitful
interaction of SETI with other sectors of the astrobiological
enterprise; this pragmatic argument has been developed in more
details in \'Cirkovi\'c (2012). It is even more important to
emphasize such integrative concepts in an epoch in which SETI
suffers from serious perception and image problems. One such
novel and integrative concept has suggested in Bradbury et al.
(2011):
}

\end{multicols}

\vfill\eject

\begin{multicols}{2}
{

\mc{\textit{Four strategies that characterize our supplemental approach,
what we have dubbed Dysonian SETI (...):}

\vskip1mm

\noindent\textit{1. The search for technological products, artefacts and
signatures of advanced technological civilizations.}

\vskip1mm

\noindent\textit{2. The study of postbiological and artificially
superintelligent evolutionary trajectories, as well as other relevant fields
of future studies.}

\vskip1mm

\noindent\textit{3. The expansion of admissible SETI target spectrum.}

\vskip1mm

\noindent\textit{4. The further development and study of astrobiology and the
achievement of tighter interdisciplinary contact with related
astrobiological subfields, including magisteria like computer science,
evolutionary biology, etc.}}

This leads to an entire new game. We are witnessing renaissance
of the \textit{extragalactic }SETI searches, most notably the \^G
infrared search for Type 2.x/Type 3 civilizations (Wright et al.
2014a, b, Griffith et al. $\!$ 2015) and the search for
stellar-powered Type 3 civilizations by using the Tully-Fisher
relation (pioneered by Annis 1999; for new attempts see
Calissendorff 2013, Zackrisson et al. 2015). Both these original
and dynamical approaches share the grounding directly inspired by
Kardashev's classification and the Dysonian SETI. There is reason
to believe, therefore, that the extent of SETI activities will
increase and diversify in the near future, so the present topic
will become more and more relevant in the years ahead.

While it was obvious at the time of its origin, subsequent use
(and occasional misuse) of Kardashev's scale has obscured the key
fact: it was meant to represent a practical guideline for what
could be expected in the course of SETI searches, not a profound
theoretical insight into the nature of extraterrestrial
intelligence. In other words, a rule-of-thumb good mason need
before starting work on any building. As Kardashev modestly put
it:\footnote{Kardashev (1964), p. 221.}

\mc{\textit{[W]e should like to note that the estimates arrived at here
are unquestionably of no more than a tentative nature. But all of them bear
witness to the fact that, if terrestrial civilization is not a unique
phenomenon in the entire universe, then the possibility of establishing
contacts with other civilizations by means of present-day radio physics
capabilities is entirely realistic. At the same time, it is very difficult
to accept the notion that, of all the 10$^{11}$ stars
present in our Galaxy, only near the
sun has a civilization developed.
It is still more difficult to extend this infer-
}}

\vskip-.5cm

\mc{\textit{ence to the
10$^{10}$ galaxies existing in the portion of the universe
accessible to observation. In any case, the deciding word on this question
is left to experimental verification.}}

We shall try to show, however, that it is much more than that --
and that, with some quite natural refinements and fine-tunings,
it can reasonably hope to guide our activities in the SETI
domain, \textit{both\/} practical \textit{and\/} theoretical, for
quite some time to come (perhaps even to the centennial, if it is
not too pretentious to speculate).

The rest of this review is organized as follows.~In Section 2,
I consider the relation of Kardashev's scale and the notions of
detectability, observation-selection effects, and astrobiological
landscape, which can offer some new and provocative perspectives
on the place and role of Kardashev Types in our overall
astrobiological research. While Section 3 is devoted to
refinements such as fractional Types (notably Types 2.x, of
relevance for practical SETI), Section 4 deals with extensions
such as Type 4. The theme of linking Kardashev's scale with the
future of the universe and the future of humanity is reiterated
in Section 5, where I suggest some strategies for undermining the
applicability of the scale to practical SETI searches, and on two
scenarios demonstrate how we could obtain wrong inferences from
naively sticking to the definition of individual Types. In the
concluding section, the emphasis on detectability is reiterated
in light of the preceding analysis, some orthogonal dimensions
for quantifying advanced technological societies are suggested,
and several directions for further research are outlined. Table 1
is of particular importance as it presents the maximal
generalization of the scale obtained while retaining the
organizing principle introduced by Kardashev.

\vskip-.4cm

\section{2. DETECTABILITY, SELECTION EFFECTS, AND THE ASTROBIOLOGICAL
LANDSCAPE}

\vskip-.4cm

\textit{If you do not expect the unexpected, you will not find it; for it is hard to be
sought out and difficult.} The pronouncement of Heraclitus of Ephesus obtains a
particular flavor in
SETI studies, where proponents have often been labeled speculative fantasts
or pseudoscientists. Kardashev's classification often risks similar fate --
its targets are too often dismissed in a hand-waiving manner, without real
understanding. And yet, for the reasons given above, the need for some
taxonomical form has re-appeared time and again.

We may start with limited and modest attempts at better elucidation and/or
modification of the Kardashev's scale, most notably by Carl Sagan, and
recently by Robert Zubrin (Zubrin 1999). In 1973, Sagan wrote:\footnote{Sagan
[1973] (2000), p. 234. The 2000 edition, produced by Jerome Agel, with
contributions by Freeman Dyson, Ann Druyan, and David Morrison, as well as
fine illustrations by Sagan's old friend and collaborator Jon Lomberg among
other artists, is a very welcome testimony on the freshness and importance
of Sagan's thought for the new millenium.}
}

\end{multicols}

\vfill\eject

\begin{multicols}{2}
{

\mc{\textit{The energy gap between a Type I and a Type II civilization,
or between a Type II and a Type III civilization is enormous---a factor of
about ten billion in each instance. It seems useful, if the matter is to be
considered seriously, to have a finer degree of discrimination. I would
suggest Type 1.0 as a civilization using 10$^{16}$ watts
for insterstellar communication; Type 1.1, 10$^{17}$
watts; Type 1.2, 10$^{18}$ watts, and so on. Our present
civilization would be classed as something like Type 0.7.}}

\vskip3mm

The equivalent formula suggested by Sagan would be

\begin{equation}
\label{eq1}
n = 1 + \frac{1}{10}\log _{10} \left( {\frac{E}{10^{16}\;\mbox{W}}}
\right),
\end{equation}

\noindent where $n $ is the Kardashev type. It agrees within an
order of magnitude with Kardashev's initial estimates, and Type
2.0 corresponds to a civilization managing total energy emitted
by the Sun (1 $L_{\odot}$).

This immediately gives us a hint that the energy values in
Kardashev's scale should not be taken literally. Solar luminosity
is convenient for \textit{us}, since we are accustomed to it in
myriad ways, not only in our astronomy, where $L_{\odot}$ is a
natural unit for stellar luminosities. But majority of stars in
the Milky Way are less luminous than our Sun, and the same holds
for majority of stars with potentially habitable planets, even in
properly conservative estimates which do not consider M-dwarf
systems as habitable (e.g. Vukoti\'c 2010). If we add M-dwarfs,
according to recent rather liberal models of habitability (Heath
et al. 1999, Tarter et al. 2007) and the conclusion that a large
fraction of them possess Earth-like planets (Petigura et al.
2013), the difference between the median luminosity and
$L_{\odot}$ becomes profound. And recall that Kardashev's scale
is justified \textit{relative }to the each individual
extraterrestrial civilization -- in a sense, the \textit{value
}of a particular milestone is important only insofar it indicates
the level of complexity and the magnitude of capacities achieved
locally.

By Eq.~(\ref{eq1}), the humanity in 2012 was of Kardashev's type 0.72. Human energy
resources are about 1.77 $\times $ 10$^{13}$ W, roughly equal to the
geothermal energy production in Earth's interior (Shimizu et al. 2011) and
are still much smaller than the total energy reserves available on our
planet, especially if nuclear energy of both fission and fusion is taken
into account. It is still much smaller than the total Solar irradiance
integrated over Earth's geometric cross-section at the upper atmosphere
(about 1.74 $\times $ 10$^{17}$ W). So, there is much more room for growth.
On the other hand, the increase in power consumption of human civilization
has been exponential, at least during the last two centuries, so any
reasonable projection at timescales negligible in astrophysical terms will
lead us very soon to the Type 1 and subsequently -- barring a global
catastrophe -- to the 1.x status.

The key emphasis implied by Kardashev's scale -- and the reason,
I suspect, for its apparent longevity -- is the focus on
\textit{detectability\/} of a technological civilization. Although
it is often downplayed in the historical SETI discourse,
detectability is clearly the most important parameter in
\textit{both} theoretical and practical SETI research (Tarter
2001, Duric and Field 2003) -- and the one which is clearly very
difficult to quantify in any detail. Kardashev's scale gives a
very crude, but still quite functional, way of quantifying the
\textit{possible detectability} of a technological civilization.
In other words, it gives us a benchmark for gauging and comparing
entities in an entirely new and previously unknown context;
benchmarks which are at least as useful as those in computer
science or risk analysis.

This can be understood in the following manner. Morphology of any
biosphere \textit{not} including technological civilization is
entirely product of biological evolution (presumably following
upon prebiotic chemical evolution). Therefore, such morphologies
are located within the huge parameter space of biological
evolutionary processes -- the ``library of Mendel'' in Dennett's
(1995) famous metaphor. With the advent of intelligent observers
and their culture, including technology for modifying physical
and biotic environment, complexity increases tremendously and the
corresponding parameter space expands. Evolutionary trajectories
now lead to many more options and the number of corresponding
morphologies exponentially increases in the course of the
cultural evolution. Navigating this gigantic parameter space in
search for something which could be detected by our meager SETI
capabilities is completely hopeless {\it unless we find some
simple way of coarse-graining it\/}; some crude way of interposing
partitions and ordering this myriads of possibilities around a
simple handle. This handle is power management/consumption and
this task was fulfilled by Kardashev's scale.

The cardinal virtue of Kardashev's particular scheme -- and one which has
not been entirely appreciated within the framework of the orthodox SETI so
far, remaining an active challenge -- is that it exactly speaks in the
\textit{language of detectability}. It sets the limits to what is achievable,
within the known laws of
physics, and it is exactly those limits which need to be probed, especially
in light of the fact that we now know that some of the Earth-like planets
are multiple Gyr older than the Earth. According to the results of
Lineweaver (2001), as well as Lineweaver et al. (2004), the
median age of the Earth-like planets is:

\vskip-3mm

\begin{equation}
\label{eq2}
\tau _{\mathrm{med}}  = \left( {6.4\pm 0.9} \right)\times 10^9\quad
\mbox{yrs},
\end{equation}

\noindent
which strongly suggests that Copernicanism is correct, at least regarding
the ages of potential biospheres in the Galaxy (Lineweaver and Davis 2002,
\'Cirkovi\'c 2009, 2012). In the same time, the question why we do not (yet)
perceive any traces, manifestations, or other sorts of evidence of Gyr-older
civilizations becomes particularly pertinent, since naive Copernicanism
would suggest that the median age of \textit{technological civilizations }is
correspondingly larger than the case
on Earth. And, as usually shown in discussions of Fermi's paradox, the
difference between timescale in Eq.~(\ref{eq2}) and the age of our Earth and the Solar
System is more than enough not only to colonize the Galaxy, but presumably
also to create a Type 3 civilization.

Obviously, \textit{there is no Type 3 civilization in the Milky
Way at present}. This basic empirical fact is worth emphasizing
for several reasons, notably because it is one way of formulating
the familiar Fermi's paradox (we would naively expect one to form
on known temporal and spatial scales, or at least to witness some
of the manifestations of its emerging).\footnote{Of course, there
are some weird possibilities which should be mentioned for the
sake of completeness. For instance, one of the hypotheses
suggested by Olum (2004) to account for some strange consequences
of the anthropic reasoning is that we are actually part of a
larger galactic civilization or a ``lost colony'', without being
aware of the fact. } In addition, we have good reasons to assume
that there has been no Type 3 civilization \textit{in the past}
of our Galaxy. We cannot be entirely certain on this, but
considering the fact that we see no traces of Solar System being
ever colonized by the galaxy-spanning large civilization, nor any
artifacts or traces of such a civilization, we can exclude this
possibility with reasonably high degree of confidence. This
hinges crucially on the definition of Type 3 civilizations.
Smaller civilizations (of what I shall call Type 2.x, see Section
3 below) could certainly exist in Milky Way's past -- whether any
exist at present is an interesting problem in SETI studies.

But by both original Kardashev's and the modified Zubrin's
understanding, there is simply no possibility to reconcile the
existence of the local (= Milky Way-based in the further text)
Type 3 civilization with the astronomical data.\footnote{If
somebody still doubts that, let us mention in passing just a
handful out of literally hundreds of pieces of empirical evidence
for that: the star-formation rate in the Milky Way is similar to
that in other normal large spiral galaxies and to our theoretical
understanding of the process (Mutch et al. 2011); stellar
population colors (Strateva et al. 2001) or dynamical
characteristics of its disk are completely in accordance with the
models suggested for spiral galaxies \textit{per se} (Kannappan
et al. 2002). Even those wildly speculative suggestions about
possible explanations of \textit{some} astronomical anomalies by
astroengineering (e.g. Beech 1990) are squarely local; it would
not be an overstatement to claim that the (present-day) existence
of Type 3 civilization in the Milky Way is at variance with the
entire edifice of contemporary stellar and galactic astronomy.
This same evidence weights, somewhat less conclusively, but still
overwhelmingly, against the existence of a \textit{past\/} Type 3
civilization as well. The latter hypothesis compounds the problem
by postulating \textit{extinction }of a galactic-size
civilization.} So, this is another way of formulating Fermi's
paradox or the ``Great Silence'' problem (Brin 1983, \'Cirkovi\'c
2009, Webb 2015). ``Being stealthy'' is at best in tension with
the ascent along Kardashev's ladder; Type 3 civilizations could
arguably be detected over huge distances, and only by making
things extremely contrived could one conceive of a ``stealthy''
Type 3 civilization. The same persistence applies to a vanished
Type 3 civilization as well, or even more: since staying stealthy
with so large energy consumption would require much intentional
effort, it might well be the case that untended artifacts or
traces of activity of a galaxy-spanning civilization would be
detectable for a long (cosmological) time after the civilization
itself goes extinct.\footnote{Completely neglecting a very
germane problem of what could be a conceivable reason for
extinction of a galaxy-wide species (while retaining local
habitability, as testified by our existence). Such literally
cosmological agency seems to be beyond our current imagination;
see, however, Egan (1997), Reynolds (2008).} The complete absence
of such artefacts or activities discovered by astronomers in the
Milky Way thus far offer support to our working conclusion that
there was no Type 3 civilization at any point in our Galaxy's
history. We shall see later how this conclusion can be
generalized and to what important constraints in the parameter
space it points.

Whether there are Type 3 civilizations in some of the other
galaxies within our cosmological horizon remains unknown, but
some preliminary indication is that they are at best rare (Annis
1999). There is a large field for possible empirical studies
there, in searching for possible outliers in the regularity of
``natural'' properties of external galaxies, like the
Tully-Fisher relation. However, this type of empirical work needs
to be based on the further theoretical insight and possible
numerical modeling of what we can reasonably expect Type 3
civilizations to look like. This has not been done so far, for
reasons probably having to do more with conservatism and bad
public image of SETI studies in the scientific community than
with modesty.

I shall argue below that even if there is a law-like regularity
preventing the emergence of Type 3 civilizations in our past
light cone, the concept itself is still quite fruitful. In
particular it is due to the circumstance that such items in the
classification obtain truly universal value in the cosmological
context, where timescales for propagation of signals are
comparable to both astrophysical and biological evolutionary
timescales. This circumstance links large-scale properties of our
universe -- especially homogeneity as understood by the classical
cosmological principle of Eddington, Milne and other early
cosmologists -- with the concept of astrobiological evolution.
And, of course, in astrobiology we wish to obtain as universal
(and ``timeless'') perspective as possible: if humanity is, in
due course, to become a pangalactic civilization on the timescale
estimated by Fermi, Hart, or Tipler in connection with Fermi's
paradox, it is useful to have such a scenario encompassed in a
natural way into our classification. It is exactly the concern
for \textit{future\/} astrobiological evolution which prompts a
generalization of the classification to include possible Type 4
civilizations, as described in Section 3 below.
}

\end{multicols}

\vfill\eject

\begin{multicols}{2}
{

Recently, we have proposed a concept of \textit{astrobiological
 landscape\/} as a useful way of thinking about biological
evolution in the most general cosmological context (\'Cirkovi\'c
2012, \'Cirkovi\'c and Vukoti\'c 2013; for a similar concept see
also Ashworth 2014). This can be imagined as a (hyper)surface
consisting of all evolutionary trajectories starting with the
``dead space'' taking up most of the universe's volume (e.g. the
intergalactic space) and the epochs before any abiogenesis took
place anywhere. Subsequent emergence and evolution of life, as
well as the appearance of intelligence (noogenesis) represented a
particular manifestations of the increase in complexity as
functions of the spatio-temporal coordinates and probably other,
yet poorly understood, physical, chemical, etc., parameters which
drive the changes in complexity. Only a part of this vast set of
possible evolutionary trajectories will be realized within our
cosmological horizon, in analogy with the fact that only a small
subset of all evolutionary possible forms will be realized in the
actual course of the biological evolution on Earth. But in each
case, we do not have reason to suspect the existence of the vast
overarching set of consistent possibilities (astrobiological
landscape or the morphological space of evolutionary biology).

It is exactly within the landscape framework that Kardashev's
scale can be usefully interpreted as a set of attractors of
evolutionary trajectories in the high-complexity
(``intelligent'') part of the overall astrobiological landscape.
Therefore, a task for the true SETI theory, which will
undoubtedly emerge at some future point, would be to explain the
dynamics of evolving different Kardashev Type civilizations,
presumably as a function of the age of a galaxy and the domicile
planetary systems and its other astrophysical properties, in
addition to a number of chemical/biological/cognitive variables.
While this might seem a tall order at this juncture, it is not
that much different from the programme of explaining the
distribution of algae in terms of marine chemical patterns and
water temperature histories (e.g. Van den Hoek 2008), to give
just one ``mundane'' example.

\vskip-1mm

\section{3. REFINING: TYPES 2.x}

\vskip-1mm

While Kardashev's study was pioneering in linking the abstract and vague
concept of the ``level of development'' of a civilization with something as
specific and quantifiable as the energy consumption and management, it does
have some limitations which could be overcome by simple refinements. One
part of the limitations comes from the very nature of the distribution of
(baryonic) matter in the universe. While it seems clear that at least a
fraction of baryonic matter is used in building an advanced civilization, it
is not obvious that non-baryonic matter is entirely unusable in this
respect. In contrast, there are some indications that annihilation of CDM
particles and antiparticles could become a viable source of energy in
physical-eschatological future (Adams and Laughlin 1997). If advanced
technological civilizations are capable of utilizing vast amounts of
non-baryonic dark matter distributed through the halo of the Milky Way --
and other large luminous galaxies -- as either energy source or other
industrial purposes, this would not only allow much larger and more complete
control of physical environment, but also would lead to civilizational
trajectories not bound to stars and their distribution. In other words,
there could be 2.x-level civilizations not located (primarily) near luminous
stars. Only advances in fundamental physics and successful detection of CDM
particles (for a review, see Cheung et al. 2012) will throw some further
light on this part of the overall design space.

Why put an emphasis on 2.x types and not, for example, on 0.x or
1.x which are equally legitimate as fractional Kardashev's Types?
The answer is twofold. One aspect is purely practical: we might
wish to concentrate on what is most likely to be detected and yet
be viable. Suppose that we are planning a practical SETI project
and that, counterfactually, we start in the \textit{tabula rasa
}state, without any empirical information about the real universe
and, especially, the real Milky Way. What would be the best
possible target civilization, the one easiest to detect? Clearly,
that would be the Type 3 Milky Way civilization. But, clearly,
such thing does not exist at present on empirical grounds. In the
other extreme, Type 0.x civilizations are clearly empirically
allowed at our present level of knowledge and in large
quantities; indeed, it is still not possible to decisively reject
the possibility of technologically very primitive civilizations
of this sort even around the closest stars. So, there is a
see-saw situation between detectability and empirical constrains,
well-known to the more original SETI thinkers: what would be
easiest to detect is the least empirically viable, and more
empirically viable targets are proportionally harder to detect.
Such a situation implies the existence of an optimum position for
meaningful targets, the one which is easiest to detect
\textit{when all empirical constrains are taken into account}. It
is my conjecture here that this optimum lies in the Type 2.x
domain, presumably toward the lower part of the scale.

One admittedly extreme, but still instructive way of thinking
about the constraints of detectability is imposed by the
Dyson-sphere model for Kardashev's Type 2 civilizations.
Following pioneer suggestion by Dyson (1960), we could expect
that sufficiently advanced civilizations will tend to use more
and more of the ``naturally available'' nuclear fusion energy
released by its parent star. This would, to an outside observer,
look like bigger and bigger part of the stellar short-wavelength
UV and optical flux being converted into high-entropy infrared
radiation corresponding to low working temperature of the
supposed alien technology. Therefore, anomalous infrared sources
are excellent targets for the Dysonian SETI attempts and some
searches have been made in the Solar neighborhood (e.g. Jugaku et
al. 1995, Jugaku and Nishimura 2003, Timofeev et al. 2000,
Carrigan 2009); continuation of this tendency consists in already
mentioned extragalactic searches, along the lines set by Annis
(1999).

The other side of the problem deals with the measure of
detectability at each particular step (or tier) of the scale.
Whatever the \ best estimator \ of
}

\end{multicols}

\vfill\eject

\centerline{\includegraphics[width=6.00in,height=4.25in]{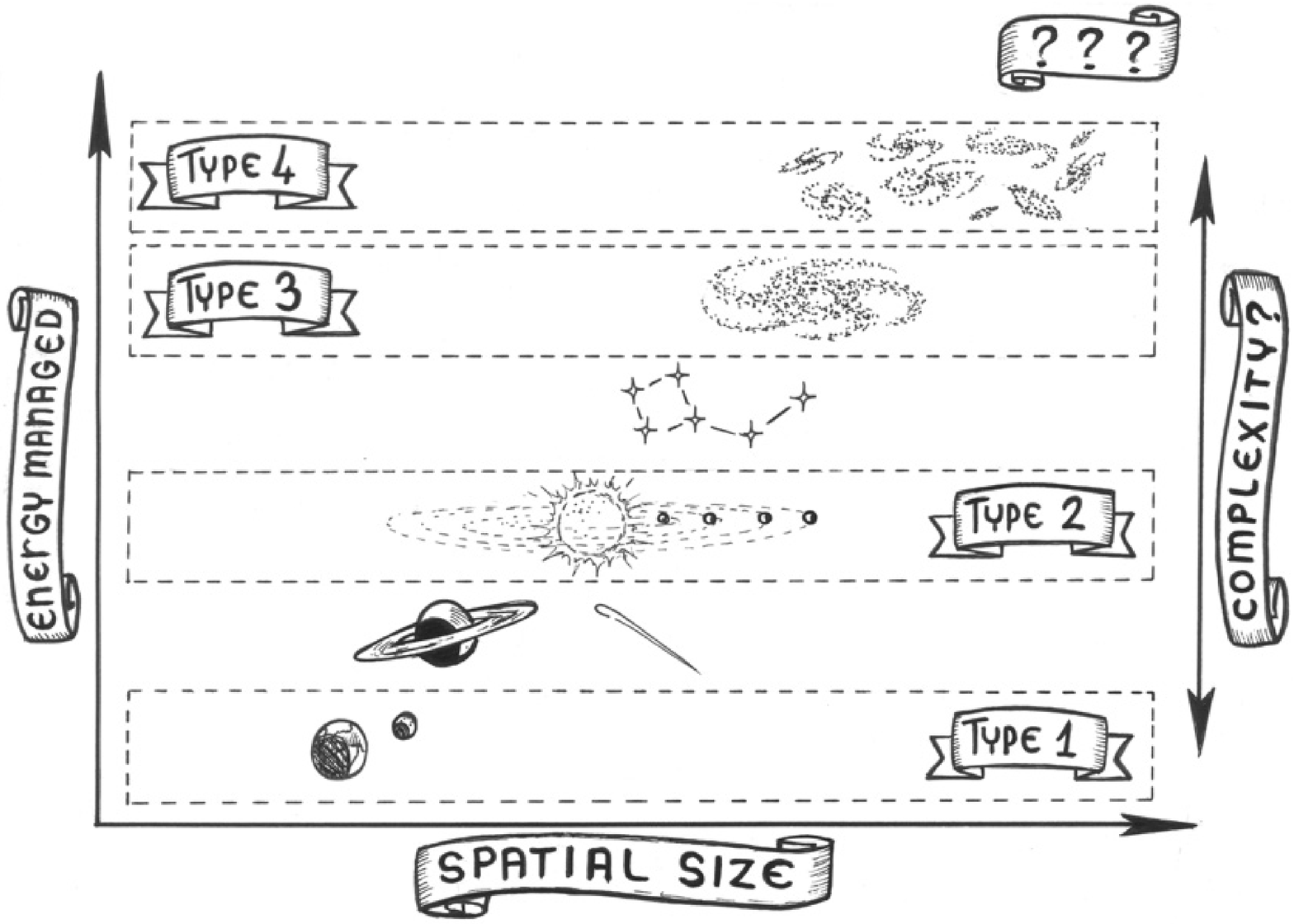}}
\label{fig2}

\vspace{-3mm}

\figurecaption{2.}{A schematic representation of Kardashev's
types in an extended framework. While the energy management is
largely determined by the fraction of cosmic resources at the
disposal of a civilization, the scale of increasing complexity is
far less certain to establish since smaller structures could, in
principle, be more complex than the large ones. Type 4 pertains to
the entire universe within our horizon -- anything above it is
necessary non-empirical and at best dependent on extremely
speculative theories. (\textit{Courtesy of Slobodan Popovi\'c
Bagi.})}

\vskip3mm

\begin{multicols}{2}
{

\noindent detectability is -- and we do not
have a clear view of this in the general case, although of course
each particular SETI activity by definition has some estimator
built in it -- we may reasonably argue that it must be
\textit{superadditive}. Namely, if $d_{m}$ is our adopted  measure of
detectability of a civilization
managing the set of resources $m $ (including the spatial volume controlled), we
may argue that it holds that, on the average:

\vskip.4cm

\begin{equation}
\label{eq3}
d_{m + n} \ge d_m + d_n .
\end{equation}

\vskip.4cm

\noindent So, for example, a civilization managing 5 planetary
systems is more detectable than either 5 civilizations managing a
single planetary system, or two civilizations, one managing 3 and
the other 2 planetary systems. Obviously, communications and
transport between the managed systems are ``extra'' potentially
detectable processes. In the worst case, those could be
undetectable, so that we would have equality in Eq.~(\ref{eq3})
above; but in realistic case we expect inequality, even very
strong one. Quantity does not make up for quality -- or a lack of
it.

Note that this conclusion does not imply that the {\it
complexity\/} measure is superadditive as well. In contrast to
most of our everyday experience, smaller structures could, from
the physical point of view, be more complex than the large ones.
In fact, the ubiquitous process of spontaneous symmetry breaking
may cause the total complexity of the whole universe to be quite
small (Tegmark 1996)! While this is admittedly an extreme view,
it still remains uncertain whether realistic civilizations, human
or extraterrestrial, possess a well-defined complexity measure
and whether such measure behaves superadditively.

It is clear what this means for practical SETI: specific searches
should make a compromise between the volume and duration surveyed
on one hand, and detectability measures on the other. Obsession
with large number of civilizations, including primitive ones
similar to humanity, evolving in parallel (large values of $N$\/
in the Drake equation) should be toned down, and traces and
manifestations of interstellar colonization and appropriate
energy consumption should be given higher priority. The spectrum
of possible targets should be increased by novel, original and
creative theoretical work, conservative hand-waiving
notwithstanding.
}

\end{multicols}

\vfill\eject

\begin{multicols}{2}
{

\section{4. EXTENSIONS AND MODIFICATION}

\vskip-3mm

While the galaxy-size civilization managing resources of several times
10$^{11}$ or more stellar objects and possibly even the CDM content may seem
as a crown of achievement of any intelligent community, this is not
necessarily so. Already Olaf Stapledon, in \textit{Star Maker }envisioned
an even greater
control over physical and cosmological environment as a possible next stage
in the universal evolution of complexity.\footnote{Stapledon (1937), esp.
chapters 9--11.} Stapledon's narrator has a vision in which he travels
through space visiting alien civilizations from the past and the future,
before finally encountering the eponymous Star Maker, an ``eternal and
absolute spirit'' who has created all these worlds in a succession of
experiments. Each experiment is a universe, and each serves designing the
next one a little better. While creation of ``basement universes'' is a
possible activity of advanced civilizations which has already been discussed
a bit by humans (e.g. Sato et al. 1982, Farhi and Guth 1987, Holt 2004),
one should keep in mind that this of itself does not guarantee further
ascent on the Kardashev's scale. Depending on the energy requirements for
such feat -- if possible within the framework of the correct theory of
quantum gravity -- it is perhaps not necessary for a civilization to be
particularly high on Kardashev's ladder to create a ``basement universe''.
Again, as in the case of possible utilizing of CDM for industrial purposes,
only further advances in fundamental physics will be able to judge to what
extent these possibilities are realistic; up to that moment, we include them
here in an extended classification (Table 1 below) as logically possible.

In the overall context of rather limited \textit{research}
interest for SETI issues, it is not surprising that the
alternatives sometime proposed have not really become comparable
in popularity in spite of being potentially more realistic and
useful. For instance, Kecskes (1998, 2009) proposes a complex
level-based hierarchy of civilizations ordered in terms of
transport, communications, material, and energy resources. It
contains four basic types (``planet dwellers, asteroid dwellers,
interstellar travellers, interstellar space dwellers''), to which
six additional types have been subsequently added. The motivation
is to avoid some the pitfalls and weaknesses of Kardashev's
scale, notably, what Kecskes and many others have perceived as
``bigger is better'' error. While the motivation is sound (and
can be illustrated by examples of defeating scenarios; see
Section 5 below), the resulting scheme is too complicated to gain
wide acceptance any time soon. The discussion of what Kecskes
calls measure of ``advancedness'' clearly shows, however, that
much multidisciplinary research, including that in social sciences
and economy, is necessary to achieve anything more realistic and
fine-grained than what Kardashev's crude scheme offers.

Gal\'{a}ntai (2004) suggests adopting designation Type 4 for the
natural extension of previous Kardashev's types, but presents
strong philosophical arguments why we should not hope to detect
it at all, even if it exists; some of those were prefigured in a
superb story-essay of Stanislaw Lem entitled ``New Cosmogony''
(Lem 1993; the original was written in 1971). Note that Lem's
conclusion is more optimistic in this respect, since he
speculates that it is exactly the influence of universe-wide
civilizations \textit{on effective local laws of physics} which
could give evidence of their existence, while the ``classical''
scenario of detection or communication is, of course, excluded.
In a subsequent work, Gal\'{a}ntai (2007) rejects Kardashev's
classification entirely and argues that a taxonomy more
appropriate for both SETI and future studies would be one based
on robustness toward various catastrophic risks. So, from level I
where a puny local disaster can destroy civilization, to level V
where potentially immortal civilizations, immune to all kinds of
threats, could be found, we have a wide spectrum of
possibilities. Gal\'{a}ntai's scheme is open to criticism that
robustness of this kind is rather unlikely to be detected from
afar; in addition, the author is forced to admit that higher
levels of robustness are connected to wider spatial range, i.e.
to the same process of interstellar colonization and resource
utilization on which Kardashev's scale is based. In a sense, this
is an unavoidable compromise Gal\'{a}ntai -- together with other
Kecskes, Zubrin, and other authors -- is forced to make with
Kardashev, if the proposed taxonomy is to retain a degree of
relevance for practical SETI projects.

We should also consider those aspects of cultural evolution leading to the
inward bounds. British astrophysicist John D. Barrow suggests adding a
number of civilization types based on the level of control of the
\textit{micro}world:\footnote{Barrow (1999), p. 133.}

\mc{\textit{Type I-minus is capable of manipulating objects over the
scale of themselves: building structures, mining, joining and breaking
solids;}

\vskip1mm

\noindent\textit{Type II-minus is capable of manipulating genes and altering
the development of living things, transplanting or replacing parts of
themselves, reading and engineering their genetic code;}

\vskip1mm

\noindent\textit{Type III-minus is capable of manipulating molecules and
molecular bonds, creating new materials;}

\vskip1mm

\noindent\textit{Type IV-minus is capable of manipulating individual atoms,
creating nanotechnologies on the atomic scale and creating complex forms of
artificial life;}

\vskip1mm

\noindent\textit{Type V-minus is capable of manipulating the atomic nucleus
and engineering the nucleons that compose it;}

\vskip1mm

\noindent\textit{Type VI-minus is capable of manipulating the most elementary
particles of matter (quarks and leptons) to create organized complexity
among populations of elementary particles;}}

%\vskip2mm

\mc{\textit{culminating in:}

\vskip2mm

\noindent\textit{Type $\Omega $-minus is capable of manipulating the basic
structure of space and time.}}

On this scale, the present-day humanity is about Type II-minus, with some
aspirations toward Type III-minus in fields like graphene or nanotube
research. Obviously, levels of Barrow's scale are not entirely decoupled
from Kardashev's Types since, for example, it is only reasonable to assume
that Barrow's Type V-minus civilizations could achieve nuclear fusion in
more efficient way than just to utilize natural stellar fusion via
Dyson-shell-type constructions. Possible or likely
interrelations between
the two scales present a fascinating topic for exploratory
engineering (Inoue and Yokoo 2011, Vidal 2014), which lies beyond the
scope of the present review.

\section{5. SOME DEFEATER SCENARIOS: INTRODUS AND THE KRELL MACHINE}

\vskip-1mm

In order to facilitate further discussion of limitations of the
Kardashev scale, I find it useful to consider possible scenarios
of systematically erroneous perception of an advanced
technological civilization, escaping the clauses of the taxonomy.
Those defeater or outlier scenarios might be of immense interest,
since their study could, at least in principle, provide deeper,
``low-level'' explanations for the regularities underlying
Kardashev's scale. As is usual in science, we arrive to
satisfactory insight into law-like regularities by investigating
counterfactual cases and exceptions from the rule. The fact that
inspiration for these is found in pop-cultural references should
not be cause for any greater hesitation in considering them
compared to the case of conceiving them as pure thought
experiments. On the contrary, examples from literary or
cinematographic fiction are advantageous over the pure thought
experiments, since often develop significant details (thus
guarding us from the coherence gap problem, see Havel 1999). It is
also likelier that they will provoke a serious debate on the
structure of the problem and possible resolutions.

I have labeled these scenarios \textit{defeating scenarios }(or
defeaters), since they represent cases of dissonance between the
real states of affairs and our practical prospects of detection
in SETI projects, while led by Kardashev's scale. In the sense
emphasized in the introductory section Kardashev's scale is
regarded as a practical tool for thinking and searching; if the
reality is dramatically different from any cases included in an
anyway flexible and wide framework, then the framework is
essentially defeated from the practical point of view (while it
might retain some theoretical relevance). While necessarily
subjective, these two scenarios -- or two bunches of scenarios
--encapsulate major points which need further elaboration in more
detailed, quantitative models.

\textbf{Introdus}: An advanced technological civilization might
consist of individuals uploaded into virtual reality supported by
extremely miniaturized and energy-efficient pieces of hardware
containing a finite number of individuals sharing, at the basic
level, the same virtual reality (``polises''). Such pieces of
hardware are very small -- in comparison to astronomical and even
geological length scales --and might be undetectable directly
over interstellar distances even if no stealth is desired or
implemented. In terms closer to the spirit of Kardashev's scale,
their efficiency might be so high that no significant amount of
waste heat and other products could be detectable above the
natural background. Since a particular ideal of ``perfection''
has been achieved, no significant interaction with the physical
surrounding is necessary -- and it might even be regarded as
culturally or morally undesirable.

Of course, the Introdus scenario is inspired by Greg Egan's
polises in his imaginative and intriguing 1997 novel
\textit{Diaspora}. It is also related to concepts such as John
Smart's ``transcension'' (Smart 2012) or Kurzweil's
``singularity'' (Kurzweil 2005, Chalmers 2010), although these
concepts possess other important elements which are irrelevant
from the point of view of SETI. The Introdus is characterized by
extremely low detectability of a civilization -- or an assembly
of civilizations -- which nevertheless possesses advanced
scientific knowledge and \textit{capacity }to perform large feats
of astro-engineering if so desired. That capacity remains, as far
as Egan's \textit{fictional }example is concerned, undeployed
since there is no impulse, or imperative to do so; even in the
face of external catastrophe which prompts the eponymous exodus
or ``diaspora'', posthuman polises remain stealthy and
undetectable. Their cross-section for any actual or even
conceivable detection and/or communication technology are
extremely low -- in dramatic contrast to their \textit{actual
level }of civilizational advancement. This could be generalized
to a wider class of evolutionary trajectories in which
optimization of research, technology and daily life becomes the
stable end-value (\'Cirkovi\'c and Bradbury 2006).

\textbf{The Krell Machine}: A complementary scenario is the one
in which large physical power is actually deployed but for some
reason is hidden or remains undetected. In the classical
science-fiction movie \textit{Forbidden Planet }(Wilcox 1956),
humans become aware of the extinct race of advanced beings of the
planet Altair IV, only known as the ``Krell''. The Krell had
reached a stage of technological and scientific development so
advanced that they were able to construct a vast underground
machine -- about 33,000 km$^{3}$ of operational volume -- with
virtually unlimited power; a machine that could turn thoughts
into reality and project that reality anywhere on the planet. The
energy resources of the Krell machine make them safely in the
Type 1.x if not Type 2 domain. (In the same time, the Krell
perfected cognitive enhancement to the level that their devices,
one of which is shown in the movie, were able to increase
intelligence and impart knowledge to any sentient being in a --
relatively -- non-invasive manner.) But they have vanished, a
rather modest interval of 200,000 years ago, leaving only the
artefacts and some terrible secrets, upon which reckless human
protagonists unwittingly stumble. As a likely a nod to Plato's
tale of Atlantis, all above-ground evidence of their civilization
has been obliterated. By its particular form and location, the
Krell machine in the movie is undetectable from afar -- but this
might not be, and indeed is not expected to be, the essential
trait of such mega-engineering feats.

These defeaters are very different in details, but share two key
common properties: (i) their detection cross-section is low; and
(ii) they can simulate -- poorly, perhaps -- multiple Kardashev's
types at once. From the point of practical searches, they are
likely to lead to confusion, since they are both hard to detect
at interstellar distances, and even if detected, are likely to be
\textit{misclassified}. (In Wilcox's movie, such
misinterpretation is one of the generators -- pun intended! -- of
the dramatic plot.)

The concept of \textit{galactic/interstellar archaeology
}(Campbell 2006, Carrigan 2012, Davies 2012) is another important
idea following not only from Kardashev's classification, but also
from the entire unconventional, Dysonian thinking about detecting
traces and manifestations. Just as terrestrial archaeology uses
artefacts of ancient civilizations to uncover their existence and
properties, so interstellar archaeology hopes to do the same for
artefacts of extraterrestrial civilizations. It's a ``parallel
track for SETI'' in words of Paul
Gilster,\footnote{{http://www.centauri-dreams.org/?p=11237.}}
 clearly affirming the relevance of Kardashev's thought.

\vskip-1mm

\section{6. DISCUSSION: DETECTABILITY, RATHER THAN DETECTION}

\vskip-2mm

In spite of all limitations and criticisms in the last 50 years,
Kardashev's scale remains the most popular and cited tool for
thinking about advanced extraterrestrial civilizations. Its
one-parameter nature has been regarded, seemingly paradoxically,
as both its strength and weakness. The analysis given above and
summarized in both Tables 1 and 2, strongly suggests that
one-parameter scale is still very much sufficient to both (i)
delineate vast domains of our ignorance, and (ii) help formulate
research programs and explanatory hypotheses aimed at diminishing
that ignorance. Exceptional cases -- i.e. evolutionary
trajectories of advanced civilizations leading them off the
Kardashev scale -- are still too rare, bizarre, and seemingly of
small practical import; those exceptions are, of course, fully
deserving of further study, especially through numerical models
and simulations. Coupled with a convenient scale describing the
dominion over microscopic physics (chemistry, biology), such as
the Barrow scale mentioned above, Kardashev's scale remains the
best taxonomical tool for both astrobiology/SETI and future
studies.

Thus, in stark contrast to most of the other hot topics of early SETI days
(like the Drake equation), Kardashev's classification aged surprisingly
well. There is not that many concepts in astronomy and related sciences
which are still active and used after more than half a century -- and in the
case of Kardashev's scale the usage seems to be more frequent in recent
years. For instance, even cursory search at the NASA ADS database shows that
more than a half of publications mentioning Kardashev's scale in the title
or the abstract have been published since the beginning of this century.

Near the very end of his original paper, Kardashev offered the
following assessment: ``The discovery of even the very simplest
organisms, on Mars for instance, would greatly increase the
probability that many Type II civilizations exist in the
Galaxy.'' This shows a rather far-reaching awareness of the
issues which will much later become part of the ``rare Earth
hypothesis'' of Ward and Brownlee (2000), as well as most of the
mainstream astrobiological thinking of today (Chyba and Hand
2005). Universality of evolution, with many peaks of complexity
in the astrobiological landscape, leads hierarchically to
different fruits in different locales, all having place within
the same huge morphological space. The most complex parts of the
astrobiological landscape, corresponding to advanced
technological civilizations, will open quite new design spaces.
In practical terms, reasoning upon which Kardashev's conclusion
is based serves as a prescient introduction into the
\textit{exploratory engineering }(cf. Armstrong and Sandberg
2013) -- we could:

\mc{$\bullet$ ask ourselves what kind of technologies is required for each step on the
scale;

\vskip2mm

\noindent $\bullet$ construct a research program to outline the necessary resources and skills
for each particular item; and

\vskip2mm

\noindent $\bullet$ determine the optimal method of detection and estimate the magnitude of
detectability, relative to the natural ``noise''.}

These steps show how the discussion about SETI is deeply
connected with both considerations of engineering and cultural
evolution on one hand, and observational astronomy
(detectability) on the other. We can even go some steps further
and consider epistemological and even ethical consequences
following from the discovery of possible extraterrestrial
intelligent artefacts, with all implications of a long-term
planning, stable society. Recent controversy over the lack of
flux from KIC 8462852 (Boyajian et al. 2015, Marengo et al. 2015,
Wright et al. 2015) is just one instance of the possible
formulation of explanatory hypotheses directly motivated by
Kardashev's scale and its ramifications. It is the prediction
following from the overall framework of detectability, Dysonian
SETI and the logic of Kardashev's scale that the number of such
hard cases in which purely ``natural'' (i.e. non-intentional)
explanations are progressively harder and harder to find will
increase with the number and sensitivity of our
}

\end{multicols}

\vfill\eject

\begin{multicols}{2}
{

\noindent detectors, in
both intragalactic and extragalactic domain.\footnote{This might
have an interesting consequence for the concept of ``success'' or
``discovery'' in the domain of SETI studies. In contrast to the
conventional image of ``first contact'' powerfully suggested by
the pop-cultural discourse (e.g. Sagan (1985) and the subsequent
movie), supported by the orthodox SETI circles, especially
radioastronomers (Tarter 2001), and encoded in the famous
``Wow!'' signal (e.g. Gray and Marvel 2001), we might not have
any particular decisive moment of discovery. Rather, we might face
slow accruement of ``inexplicable'' cases without natural or
non-artificial explanation, leading gradually to mainstream
acceptance of astroengineering as not only legitimate, but the
best explanation.}

A proposed generalization of Kardashev's classification may be
schematically presented in the Table 1. While rather conservative
in comparison to the overhauls suggested by Kecskes, Vidal, or
Gal\'{a}ntai, it still formalizes the expansion of thinking
prompted by the astrobiological revolution and the Dysonian SETI.
While one could criticize the emphasis on power consumption and
``bigger is better'' thinking inherent in it, it is still more
amazing how few exceptions or defeaters have actually been
conceived in the literature (both discoursive and fictional) so
far. Some of them are summarized in Table 2.

The emphasis on \textit{detectability} is, as justified above, a
particularly salient feature of Kardashev's and
Kardashev-inspired schemes and the one which still needs to be
tirelessly repeated, more than a half century later. Namely, it
goes against the grain of the orthodox SETI with its blind
insistence on large values of $N $ in the Drake equation as a
good predictor of the success of practical search activities. But
the value of $N$\/ is, to a large extent, a red herring. It does
not require a sophisticated analysis to conclude that a
\textit{single} ($N=1$) Type 3 civilization is a better SETI
target than a hundred or a thousand or perhaps a million of
humanity-level Type $< 1$ civilizations. ``Better'' here means
easier to detect signal and easier to recognize its artificial
nature. This simple insight has in the meantime been
observationally operationalized and used in very real SETI
surveys of external galaxies (Annis 1999, Wright et al. 2014a,b).
Taking into account superadditivity, as discussed above, leads to
similarly obvious conclusions when civilizations of the 2.x and
even 1.x Types are concerned. Since the amount of resources, like
the energy whose fraction is directly or indirectly used for
emitting those signals potentially detectable by other observers,
is huge but definitely finite, Kardashev's ladder also tells us
simple, but important truth that the naive idea about ``place for
everybody and everything'' in the vastness of the universe is, in
fact, wrong.
}

\end{multicols}

\vskip2mm

\centerline{\textbf{Table 1.} An extended view of Kardashev's scale.}

\vskip2mm

\noindent\begin{tabular}
{|p{100pt}|p{180pt}|p{150pt}|}
\hline
\textbf{Kardashev's Type}&
\textbf{manages resources}&
\textbf{comments} \\
\hline \textbf{0} \par \ \ \textbf{-- 0.x}& pre-technological
society \par \ \ -- of particular area of the planet, or a \par
particular type of planetary resources&
humanity about 0.8 at present \\
\hline
\textbf{1} \par \ \ \textbf{-- 1.x}&
of its home planet \par \ \ -- of a number of planets and other \par
\ \ \ \ bodies within a planetary system &
Introdus / Krell machine type \par scenarios as exceptions \par \ \ -- detectability \textit{superadditive} \\
\hline
\textbf{2} \par \ \ \textbf{-- 2.x}&
of its home star and planetary system \par \ \ -- of a number of planetary systems \par
\ \ \ \ within a region of the home galaxy&
Dyson shell-like contraptions \par \ \ -- detectability \textit{superadditive} \\
\hline
\textbf{3} \par \ \ \textbf{-- 3.x}&
of its home galaxy \par \ \ -- of a number of galaxies within a \par
\ \ \ \ region of the universe&
absent from the Milky Way, \par closer galaxies \par \ \ -- detectability \textit{superadditive} \\
\hline
\textbf{4} \par \ \par \ \ \textbf{-- 4.x}&
of the universe within cosmological \par horizon \par \ \ -- of a number of topologically \par
\ \ \ \ connected universes&
causal disconnect occurs at \par particular epoch, depending on \par the cosmological model \par
\ \ -- detectability irelevant?? \\
\hline
\textbf{5}&
of the multiverse&
topological structure crucial \\
\hline
\end{tabular}
\label{tab1}

\vskip.7cm

\noindent\textbf{Table 2.} As an addition to Table 1, here I list conceivable
ways in which Kardashev's scale could be considered incomplete. While these
defeater scenarios are arguably too speculative or of too small probability
measure, we should keep them in mind in surveying the overall
astrobiological landscape.

\vskip2mm

\noindent\hskip3mm\begin{tabular}
{|p{167pt}|p{264pt}|}
\hline
\textbf{Kind of incompleteness}&
\textbf{defeater scenarios} \\
\hline
\textbf{BEFORE THE SCALE}&
dead space,  ``rare Earth'' \\
\hline
\textbf{SCALE IMPRACTICAL}&
galactic archaeology, Introdus, the Krell Machine \\
\hline
\textbf{BEYOND THE SCALE}&
simulated universes, ``new cosmogony'', Boltzmann brains \\
\hline
\end{tabular}
\label{tab2}

\vskip.5cm

\begin{multicols}{2}
{

While this obvious conclusion has occasionally been recognized,
its importance and practical consequences for SETI projects have
not been fully understood and adopted so far. Instead, rather
unhealthy obsession with the value of Drake's $N$\/ continues to
this day. Rethinking Kardashev's classification should have a
salutary effect in this area as well. The truly important issue,
especially following the null result of the \^G search for Type 3
civilizations (Griffiths et al. 2015), the most detailed and
comprehensive such observation effort thus far, is whether Type
2.x civilizations are detectable from a range of realistic
distances, intragalactic as well as intergalactic. This challenge
for innovative, imaginative, creative, and bold SETI will remain
open for at least a couple of decades to come. Among other
things, it will help understanding the prospects and pitfalls of
the future of humanity itself, hopefully contributing to a new
ecological and ethical consensus necessary for the long-term
survival and prosperity of our species.

\acknowledgements{I wish to thank Jelena Dimitrijevi\'c, Anders
Sandberg, Stuart Armstrong, Branislav Vukoti\'c, Nick Bostrom, Slobodan
Popovi\'c, Slobodan Perovi\'c, Ivana Kojadinovi\'c, Karl Schroeder, Petar Gruji\'c,
Jelena Andreji\'c, Mom\v{c}ilo Jovanovi\'c, Goran Milovanovi\'c, Eva Kamerer, Du\v{s}an
Indji\'c, Zona Kosti\'c, George Dvorsky, Zoran Kne\v{z}evi\'c, Steven J. Dick,
Jacob Haqq-Misra, the late Robert Bradbury, and the late Branislav
\v{S}impraga for many pleasant and useful discussions on the topics related
to the subject matter of this study. Aleksandar Obradovi\'c, Du\v{s}an
Pavlovi\'c, and Seth Baum kindly helped in obtaining some of the crucial
references. This is also an opportunity to thank KoBSON Consortium of
Serbian libraries, NASA Astrophysics Data System and incredibly useful
websites {{http://arxiv.org/}} and
{{http://tvtropes.org/}}.~The author has been supported by the Ministry of
Education, Science and Technological Development of the Republic of Serbia
through grant ON176021.}

\vskip.7cm

\references

Adams, F. C. and Laughlin, G.: 1997, \journal{Rev. Mod. Phys.}, \vol{69}, 337.

Annis, J.: 1999, \journal{J. Brit. Interplanet. Soc.}, \vol{52}, 33.

Armstrong, S. and Sandberg, A.: 2013, \journal{Acta Astronaut.}, \vol{89},
1.

Ashworth, S.: 2014, \journal{J. Brit. Interplanet. Soc.},
\vol{67}, 224.

Barrow, J. D.: 1999, {Impossibility: The Limits of Science and the Science
of Limits}, Oxford University Press, Oxford.

Beech, M.: 1990, \journal{Earth Moon Planets}, \vol{49}, 177.

Bennett, J. O. and Shostak, S.: 2011, {Life in the Universe},
3rd Edition, Benjamin Cummings, San
Francisco.

Boyajian, T. S. et al.: 2016, \journal{Mon. Not. J. Astron. Soc.},
in press (arXiv:1509.03622).

{\ }

Bradbury, R. J., \'Cirkovi\'c, M. M., and Dvorsky, G.: 2011, \journal{J. Brit. Interplanet. Soc.}, \vol{64},
156.

Brin, G. D.: 1983, \journal{Q. Jl. R. Astr. Soc.}, \vol{24}, 283.

Calissendorff, P.: 2013, A Dysonian Search for Kardashev Type III
Civilisations in Spiral Galaxies, BSc Thesis, Stockholm
University.

Campbell, J. B.: 2006, in
``{Direct Imaging of Exoplanets: Science {\&} Techniques}'' Proceedings of the IAU Colloquium {\#}200, eds. C. Aime and F. Vakili,
Cambridge University Press, Cambridge, p.247.

Carrigan,~R.~A.: 2009, \journal{Astrophys. J.},
\vol{698}, 2075.

Carrigan, R. A.: 2012, \journal{Acta Astronaut.}, \vol{78},
121.

Carter, B.: 2008, \journal{Int. J. Astrobiol.}, \vol{7},
177.

Carter, B.: 2012, \journal{Int. J. Astrobiol.}, \vol{11},
3.

\'Cirkovi\'c, M. M.: 2008, \journal{Collapse}, \vol{5}, 292.

\'Cirkovi\'c, M. M.: 2009, \journal{Serb. Astron. J.}, \vol{178}, 1.

\'Cirkovi\'c, M. M.: 2012, {The Astrobiological Landscape: Philosophical
Foundations of the Study of Cosmic Life}, Cambridge University Press, Cambridge.

\'Cirkovi\'c, M. M. and Bradbury, R. J.: 2006, \journal{New Astron.}, \vol{11},
628.

\'Cirkovi\'c, M. M. and Vukoti\'c, B.: 2008, \journal{Origin of Life
and Evolution of the Biosphere}, \vol{38}, 535.

\'Cirkovi\'c, M. M. and Vukoti\'c, B.: 2013, \journal{Int. J. Astrobiology}, \vol{12}, 87.

Chalmers, D. J.: 2010, \journal{Journal of Consciousness Studies},
\vol{17}, 7.

Cheung, K., Po-Yan, T., Yue-Lin, S. T. and Tzu-Chiang, Y.: 2012,
\journal{Journal of Cosmology and Astroparticle Physics}, \vol{05}, 001.

Chyba, C. F. and Hand, K.: 2005, \journal{Annu. Rev. Astron. Astrophys.},
\vol{43}, 31.

Cocconi, G. and Morrison, P.: 1959, \journal{Nature} \vol{184}, 844.

Conway Morris, S.: 2003, \journal{Int. J. Astrobiology},
\vol{2}, 149

Conway Morris, S.: 2011, \journal{Philosophical Transactions of the Royal
Society A}, \vol{369}, 555.

Crow, M. J.: 1986, {The Extraterrestrial Life Debate 1750-1900},
Cambridge University Press, Cambridge.

Darling, D.: 2001, {Life Everywhere: The Maverick Science of Astrobiology},
Basic Books, New York.

Davies, P. C. W.: 2012, \journal{Acta Astronaut.}, \vol{73},
250.

Dennett, D.: 1995, {Darwin's Dangerous Idea: Evolution and the Meanings of Life},
Simon {\&} Schuster, New York.

Des Marais, D. J. and Walter, M. R.: 1999, \journal{Annu. Rev. Ecol. Syst.}, \vol{30},
397.

Dick, S. J.: 1996, {The Biological Universe: The Twentieth-Century Extraterrestrial
Life Debate and the Limits of Science}, Cambridge University Press, Cambridge.

Diels, H.: 1983, {Presocratic Fragments}, Naprijed, Zagreb  (in Croatian).

Duric, N. and Field, L.: 2003, \journal{Serb. Astron. J.}, \vol{167}, 1.

Dyson, F. J.: 1960, \journal{Science}, \vol{131}, 1667.

Dyson, F. J.: 2007, {A Many-Colored Glass: Reflections on the Place of Life
in the Universe}, University of Virginia Press, Charlottesville.

Egan, G.: 1997, {Diaspora}, Orion/Millennium, London.

\endreferences
}

\end{multicols}

\vfill\eject

\begin{multicols}{2}
{

\refcontinue

Ellis, G. F. R., Kirchner, U. and Stoeger, W. R.: 2004, \journal{Mon. Not. R.
Astron. Soc.}, \vol{347}, 921.

Farhi, E. and Guth, A. H.: 1987, \journal{Physics Letters B}, \vol{183}, 149.

Fogg, M. J.: 1988, \journal{J. Brit. Interplanet. Soc.},
\vol{41}, 491.

Franck, S., von Bloh, W. and Bounama, C.: 2007, \journal{Int. J. Astrobiology},
\vol{6}, 153.

Gal\'{a}ntai, Z.: 2004, \journal{Periodica Polytechnica Ser. Soc. Man. Sci.},
\vol{12}, 83.

Gal\'{a}ntai, Z.: 2007, \journal{Contact in Context}, \vol{2}(2),
{http://mono.eik.bme.hu/$\sim $galantai/articles/\break
After{\_}Kardashev{\_}Farewell{\_}to{\_}Super{\_}Civiliza\break
tons.html}.

Gilmour, I. and Sephton, M. A.: 2004, {An Introduction to Astrobiology},
Cambridge University Press, Cambridge.

Gleiser, M.: 2010, \journal{Int. J. Mod. Phys. D},
\vol{19}, 1299.

Gonzalez, G.: 2005, \journal{Origin of Life and Evolution of the Biosphere},
\vol{35}, 555.

Gonzalez, G., Brownlee, D. and Ward, P.: 2001, \journal{Icarus},
\vol{152}, 185.

Gould, S. J.: 1989, {Wonderful Life: The Burgess Shale and the Nature
of History}, W. W. Norton, New York.

Gray, R. H. and Marvel, K. B.: 2001, \journal{Astrophys. J.},
\vol{546}, 1171.

Griffith, R. L., Wright, J. T., Maldonado, J., Povich, M. S., Sigurdsson,
S. and Mullan, B.: 2015, \journal{Astrophys. J. Suppl. Series},
\vol{217}, 25.

Grinspoon, D.: 2003, {Lonely Planets: The Natural Philosophy of Alien Life},
Harper Collins, New York.

Hair,~T.~W. and~Hedman,~A.~D.: 2013, \journal{Int. J. Astrobiology},
\vol{12}, 45.

Hart, M. H.: 1975, \journal{Q. Jl. R. Astr. Soc.}, \vol{16}, 128.

Havel, I. M.: 1999, \journal{Foundations of Science}, \vol{3}, 375.

Heath, M. J., Doyle, L. R., Joshi, M. M. and Haberle, R. M.: 1999,
\journal{Origins of Life and Evolution of the Biosphere}, \vol{29}, 405.

Holt, J.: 2004, {Slate}, {http://slate.msn.com/id/2100

\noindent 715}.

Hoyle, F.: 1983, {The Intelligent Universe}, Michael Joseph Limited, London.

Impey, C. (ed.): 2010, {Talking About Life: Conversations on Astrobiology},
Cambridge University Press, Cambridge.

Inoue, M. and Yokoo, H.: 2011, \journal{J. Brit. Interplanet. Soc.}, \vol{64}, 58.

Jones, M.: 2015, \journal{Acta Astronaut.}, \vol{116},
161.

Jugaku, J., Noguchi, K. and Nishimura, S.: 1995, in ``Progress in the Search for Extraterrestrial Life,'' ed. Seth
Shostak, G., ASP Conference Series, San Francisco, p.381.

Jugaku, J. and Nishimura, S.: 2003. ``Bioastronomy 2002: Life Among the Stars'' Proceedings of IAU
Symposium {\#} 213, eds. Norris R., Stootman F.,  ASP Conference Series, San Francisco, p.437.

Kannappan, S. J., Fabricant, D. G. and Franx, M.: 2002, \journal{Astron. J.},
\vol{123}, 2358.

Kardashev, N. S.: 1964, \journal{Sov. Astron.}, \vol{8}, 217.

Kardashev, N. S.: 1997, \journal{Astrophys. Space Sci.},
\vol{252}, 25.

Kecskes, C.: 1998, \journal{J. Brit. Interplanet. Soc.}, \vol{51}, 175.

{\ }

Kecskes, C.: 2009, \journal{J. Brit. Interplanet. Soc.}, \vol{62}, 316.

Korpela, E. J., Sallmen, S. M. and Leystra Greene, D.: 2015, \journal{Astrophys.
J.}, \vol{809}, 139.

Kragh, H.: 2004, {Matter and Spirit in the Universe: Scientific and
Religious Preludes to Modern Cosmology}, Imperial College Press, London.

Kukla, A.: 2001, \journal{Stud. Hist. Phil. Sci.}, \vol{32}, 31.

Kurzweil, R.: 2005, {The Singularity Is Near: When Humans
Transcend Biology}, Viking, New York.

Lem, S.: [1971] 1993, in ``A Perfect Vacuum'',
trans. by M. Kandel, Northwestern University Press, Evanston, p.197.

Linde, A. D. and Vanchurin, V.: 2010, \journal{Phys. Rev. D},
\vol{81}, 083525.

Lineweaver, C. H.: 2001, \journal{Icarus}, \vol{151}, 307.

Lineweaver, C. H. and Davis, T. M.: 2002, \journal{Astrobiology},
\vol{2}, 293.

Lineweaver, C. H., Fenner, Y. and Gibson, B. K.: 2004, \journal{Science},
\vol{303}, 59.

Marengo, M., Hulsebus, A. and Willis, S.: 2015, \journal{Astrophys. J.},
\vol{814}, L15.

Mayr, E.: 1993, \journal{Science}, \vol{259}, 1522.

Michaud, M. A. G.: 2007, {Contact with Alien Civilizations: Our Hopes and
Fears about Encountering Extraterrestrials}, Copernicus Books, New York.

Mutch, S. J., Croton, D. J. and Poole, G. B.: 2011, \journal{Astrophys. J.},
\vol{736}, 84.

Olum, K.: 2004, \journal{Analysis}, \vol{64}, 1.

Page, D. N.: 2008, \journal{Phys. Rev. D}, \vol{78}, 023514.

Petigura, E. A., Howard, A. W. and Marcy, G. W.: 2013, {Proceedings of the
National Academy of Science}, \vol{110}, 19273.

Rescher, N.: 1985,  in ``Extraterrestrials, Science and Alien Intelligence'', ed. E. Regis, Jr.,
Cambridge University Press, Cambridge, p.83.

Reynolds, A.: 2008, {House of Suns}, Gollancz, London.

Sagan, C.: 1973, \journal{Icarus}, \vol{19}, 350.

Sagan, C.: 1985, {Contact}, Simon {\&} Schuster, New York.

Sagan, C.: [1973] 2000, {Carl Sagan's Cosmic Connection: An
Extraterrestrial Perspective}, Cambridge University Press, Cambridge.

Sagan, C. and Walker, R. G.: 1966, \journal{Astrophys. J.},
\vol{144}, 1216.

Sagan, C. and Salpeter, E. E.: 1976, \journal{Astrophys. J. Suppl. Series},
\vol{32}, 737.

Sato, K., Kodama, H., Sasaki, M. and Maeda, K.-I.: 1982, \journal{Physics Letters B},
\vol{108}, 103.

Shimizu, I. et al. (The KamLAND Collaboration): 2011, \journal{Nature Geosciences},
\vol{4}, 647.

Shklovskii, I. S. and Sagan, C.: 1966, {Intelligent Life in the
Universe}, Holden-Day, San Francisco.

Simpson, G. G.: 1964, \journal{Science}, \vol{143}, 769.

Smart, J. S.: 2012, \journal{Acta Astronaut.}, \vol{78}, 55.

Stapledon, O.: 1937, {Star Maker}, Methuen, London.

Strateva, I. et al.: 2001, \journal{Astron. J.}, \vol{122}, 1861.

Tarter, J.: 2001, \journal{Annu. Rev. Astron. Astrophys.},
\vol{39}, 511.

Tarter, J. C. et al.: 2007, \journal{Astrobiology}, \vol{7}, 30.

Tegmark, M.: 1996, \journal{Foundations of Physics Letters},
\vol{9}, 25.

Timofeev, M. Yu., Kardashev, N. S. and Promyslov, V. G.: 2000,
\journal{Acta Astronaut.}, \vol{46}, 655.

\endreferences

\vfill\eject

\refcontinue

Van den Hoek, C.: 2008, \journal{Biological Journal of the
Linnean Society}, \vol{18}, 81.

Voros, J.: 2014, in ``Teaching and Researching Big History: Exploring a New Scholarly Field'', eds.
L. Grinin, D. Baker, E. Quaedackers and A. Korotayev, Uchitel Publ. House, Volgograd, Russia, p.283.

Vidal, C.: 2014, {The Beginning and the End: The Meaning of Life in a
Cosmological Perspective}, Springer, New York.

Vukoti\'c, B. and \'Cirkovi\'c, M. M.: 2012, \journal{Origins of Life and Evolution
of Biospheres}, \vol{42}, 347.

Ward, P.: 2005, {Life as We Do Not Know It},
Viking, New York.

Ward, P. D. and Brownlee, D.: 2000, {Rare Earth: Why Complex Life Is Uncommon
in the Universe}, Springer, New York.

{\ }

{\ }

Webb, S.: 2015, {Where is Everybody? Seventy-Five Solutions to the Fermi Paradox
and the Problem of Extraterrestrial Life}, Springer, New York.

Wilcox, F. M. (director): 1956, {Forbidden Planet}, Metro-Goldwyn-Mayer Corp.

Wright, J. T., Mullan, B., Sigurdsson, S. and Povich, M. S.: 2014a,
\journal{Astrophys. J.}, \vol{792}, 26.

Wright, J. T., Griffith, R. L., Sigurdsson, S., Povich, M. S. and Mullan,
B.: 2014b, \journal{Astrophys. J.}, \vol{792},
27.

Wright, J. T., Cartier, K. M. S., Zhao, M., Jontof-Hutter, D. and Ford, E.
B.: 2015, \journal{Astrophys. J.}, in press (arXiv:1510.04606).

Zackrisson, E., Calissendorff, P., Asadi, S. and Nyholm, A.: 2015,
\journal{Astrophys. J.}, \vol{810}, 23.

Zubrin, R.: 1999, {Entering Space: Creating a Spacefaring
Civilization}, Jeremy P. Tarcher/Putnam, New York.

\endreferences

}
\end{multicols}

\end{document}